\pdfoutput=1

\documentclass[11pt]{article}
\usepackage{jheppub}

\usepackage{amsmath}
\usepackage{amssymb}
\usepackage{graphicx,color,slashed}
\usepackage{bbm} 
\usepackage{ifpdf}
\newcommand{\ft}[2]{{\frac{#1}{#2}}}

\newcommand{\cJ}{\mathcal{J}}

\newcommand{\cL}{\mathcal{L}}

\newcommand{\cN}{\mathcal{N}}

\newcommand{\cR}{\mathcal{R}}

\newcommand{\be}{\begin{equation}}
\newcommand{\ee}{\end{equation}}
\newcommand{\ba}{\begin{eqnarray}}
\newcommand{\ea}{\end{eqnarray}}

\newcommand{\N}{\mathcal{N}}

\def\rmi{{\rm i}}

\newcommand{\rf}[1]{(\ref{#1})}

\title{Nonlinear (Super)Symmetries and Amplitudes}

\author{Renata Kallosh}
\affiliation{SITP and Department of Physics, Stanford University, Stanford, CA
94305, USA}

\emailAdd{kallosh@stanford.edu}

\notoc

\abstract{There is an increasing  interest in nonlinear supersymmetries in cosmological model building. Independently, elegant  expressions for  the all-tree amplitudes in models with nonlinear symmetries, like D3 brane Dirac-Born-Infeld-Volkov-Akulov  theory, were recently discovered. Using the generalized background field method we show how, in general, nonlinear symmetries  of the action, bosonic and fermionic,  constrain amplitudes beyond soft limits. The same identities control, for example, bosonic $E_{7(7)}$ scalar sector symmetries as well as the fermionic goldstino symmetries.

We present a universal derivation of the vanishing amplitudes in the single (bosonic or fermionic) soft limit.  We explain why, universally,   the double-soft limit probes the coset space algebra. We also provide identities  describing the multiple-soft limit. We discuss loop corrections to $\cN\geq 5$ supergravity, to the D3 brane, and  the UV completion of constrained multiplets in string theory.}

\begin{document}

\maketitle

\newpage

 \tableofcontents{}

\newpage

\section{Introduction} 
The concept of a symmetry of the action was introduced into physics  by Emmy Noether \cite{Noether:1918zz}. Noether theorem relates the global symmetries of the action to the 
conservation laws. It is interesting that the actions used in the original work can have any number of derivatives and still under certain conditions specified in \cite{Noether:1918zz}  one finds a conservation of the Noether current, once the Euler-Lagrange field equations are satisfied, even when the relevant actions have an infinite number of derivatives\footnote{See \cite{Townsend:2016slt,Avery:2015rga} for a recent analysis of Noether theorems.}. Noether transformations laws acting on the fields  can be linear or nonlinear, or some combination of these. In case that Noether symmetries involve dualities and a transformation of the vector field strength's, things are more subtle and this was explained  in \cite{Gaillard:1981rj}, where the relevant  Noether-Gaillard-Zumino identities were introduced. In this paper we will not include duality symmetry involving vectors, and consider only a standard Noether type set up.

Meanwhile the development of QFT and the success of the standard model in particle physics focused the main attention to linearly realized local gauge symmetries, starting with Ward-Takahashi identities in QED in \cite{Ward:1950xp,Takahashi:1957xn} with a continued success in the case of spontaneously broken non-Abelian gauge symmetries.

Linear symmetries, where the infinitesimal transformations of the fields have a field-independent term  and a term linear in the field,  were studied intensely classically and in quantum theory.
It is well known  how to construct amplitudes with account of quantum corrections taking into account the corresponding Ward identities, following from linear  symmetries, including supersymmetries,  see for example \cite{Elvang:2015rqa}. Much less is known about the amplitudes in models with nonlinear symmetries, where the transformations of the fields has a field-independent constant term,  a term linear in the field, and also higher powers of the field, starting with terms quadratic in fields.  Global nonlinear symmetries of the action are associated with Noether theorem and it is  believed  that soft limit of amplitudes represents the nonlinear supersymmetries.

The recent interest to nonlinear symmetries, especially nonlinear supersymmetries, originate from the fact that constrained superfields describing Volkov-Akulov (VA) type models, \cite{Volkov:1972jx,Volkov:1973ix} are interesting for cosmology. The nonlinear supersymmetry transformation and a translation symmetry of  goldstino   is
\be
\delta  \lambda^\alpha =  \zeta^\alpha + \xi^\mu \partial _\mu \lambda^\alpha- i (\lambda \sigma^\rho \bar \zeta - \zeta \sigma^\rho \bar \lambda) \partial _\rho \lambda^\alpha \, .
\label{susy}\ee
The nonlinear supersymmetry transformation has a constant spinor term $\zeta$ and a term with  $\zeta$ quadratic in goldstino, as well as translation with a parameter $\xi^\mu$ which is linear in goldstino. The models of this class 
are very useful for describing dark energy and inflation, see for example, \cite{Antoniadis:2014oya,Ferrara:2014kva,Ferrara:2015tyn,Carrasco:2015pla}. In particular,  de Sitter supergravity was constructed by promoting VA supersymmetry to a local symmetry, \cite{Bergshoeff:2015tra,Hasegawa:2015bza,Kuzenko:2015yxa,Bandos:2015xnf}. A maximally supersymmetric extension of VA theory is given by a D3 brane action, and explicit d=4 Dirac-Born-Infeld-Volkov-Akulov model (DBI-VA),  is presented in  \cite{Bergshoeff:2013pia}, see Appendix A for a short review. It has 4 deformed Maxwell type supersymmetries and 4 nonlinear VA type supersymmetries. It was recently shown in \cite{Vercnocke:2016fbt,Kallosh:2016aep}, that with regard to nonlinear symmetries all these fields belong to constrained (partnerless) multiplets. This is despite the fact that at the linear level the model has Maxwell-type  ${\cal N}=4$ linear supersymmetry.

Meanwhile, an independent progress was recently made in studies of amplitudes in models with nonlinear symmetry.
The first results on amplitudes and their single and double soft limits  in models with fermionic VA nonlinear symmetries were obtained in  \cite{Chen:2014xoa}. More recently, the amplitudes were studied in 
 NLSM (nonlinear sigma models) and their supersymmetric version, VA models  and DBI-VA model \cite{Bergshoeff:2013pia}, see \cite{He:2016vfi,Cachazo:2016njl,Carrasco:2016ldy} for the most recent progress on amplitudes in these class of models. A significant role in these new constructions is played by soft theorems and recursion relations, see for example \cite{Chen:2014xoa,He:2016vfi,Cachazo:2016njl,Carrasco:2016ldy,Volovich:2015yoa,Luo:2015tat,Bianchi:2016viy} and references therein. However, it is not quite clear how exactly all newly discovered features of these amplitudes are related to nonlinear supersymmetry. For example in \cite{He:2016vfi} it was suggested  that the double-soft theorems in DBI-VA theory  will provide clues for the `mysterious nonlinearly realized (super)symmetries of the theory'.

An example of the $E_{7(7)}$ transformation on scalars in ${\cal N}=8$ supergravity is, in notation of \cite{Kallosh:2008ic}
\be
\delta y = \Sigma  +y \bar \Lambda - \Lambda y  - y\bar \Sigma y \, .
\ee
Here  $y$ is an inhomogeneous coordinate of the ${E_{7(7)}\over SU(8)}$ coset space, $\Sigma$ is a constant,  an off-diagonal part of the element of $E_{7(7)}$ and finally, $\bar \Lambda$ and $\Lambda$ represent a linear $SU(8)$ part.

In both cases the symmetry transformation has constant field-independent terms, $\zeta$ in VA case and $\Sigma$ in  $E_{7(7)}$ case, there is a linear term, translation in VA model and $SU(8)$ symmetry in $\cN=8$ supergravity, as well as  quadratic in fields, $(\lambda \sigma^\rho \bar \zeta - \zeta \sigma^\rho \bar \lambda) \partial _\rho \lambda^\alpha$  in VA case and 
$y\bar \Sigma y$ in ${\cal N}=8$ supergravity.

The purpose of this paper is to study a general  effect of nonlinear symmetry of the action on the S-matrix. 
We will generalize here  the background field method\footnote{In amplitude community the background field method in application to  gluons is known as  Berends-Giele recursion \cite{Berends:1987me}.}
and the abstract formalism of Bryce DeWitt \cite{DeWitt:1967ub} \footnote{In \cite{Honerkamp:1971sh} the  NLSM was studied in the background field formalism. A background field method for gravity including  a prediction on  1-st loop quantum corrections and absence of gauge-invariant  UV divergences  was developed in \cite{'tHooft:1974bx}. } for the case of nonlinear symmetries.
DeWitt's formalism was developed in the past mostly 
for linearly realized gauge symmetries in quantum field theories like non-Abelian Yang-Mills theory and  gravity, see for example,  \cite{Kallosh:1974yh,Grisaru:1975ei,Goroff:1985th,vandeVen:1991gw}.
A prediction for the 2-loop gravity in \cite{Kallosh:1974yh} based on background field method for the $R^3$ UV divergence was confirmed and a non-vanishing UV divergence coefficient was found in \cite{Goroff:1985th} and confirmed later still using the background field method in a different gauge in \cite{vandeVen:1991gw}.  Recently the result was confirmed in amplitude computations, using special evanescent effects in \cite{Bern:2015xsa}.
 The basic feature of  gauge theories, like the non-Abelian symmetry and gravity,  is that the off-shell local symmetry transformations remain the same with and without adding to the action  terms with higher derivatives. For example, the gauge symmetry of the non-Abelian gauge field 
\be
\delta_{\rm lin} A_\mu^a= D_\mu^{ab} \xi^b = \partial _\mu \xi^a + f^{abc} A_\mu ^b \xi^c   
\label{non-ab}\ee
is obviously linear in the field $A_\mu$. The symmetry transformation of the metric is linear in the metric $g_{\mu\nu}$, when the proper choice of variables is made  \cite{DeWitt:1967ub}
\be
 \delta_{\rm lin}  g_{\mu\nu}= D_{(\mu} \xi_{\nu)} = -g_{\mu\nu, \sigma} \xi^\sigma - g_{\nu\mu, \sigma} \xi^\sigma - g_{\mu \sigma} \xi^\sigma{}_ {,\nu}\, .
\label{grav}\ee
But the most important feature here is that {\it in both  non-Abelian gauge theory and gravity the symmetry transformations \rf{non-ab} and \rf{grav} are the same whether the classical action is used or higher derivative terms are added.} The form of the gauge symmetries 
is not affected by the changes in the action when terms are added to classical action to absorb the potential UV divergences. The structure of counterterms is defined by the gauge symmetry in the background field method, which is just a simplified form of the Ward identities \cite{DeWitt:1967ub,Honerkamp:1971sh,'tHooft:1974bx,Kallosh:1974yh,Grisaru:1975ei}.

In supergravity the off-shell local supersymmetry transformations are action independent only in case that auxiliary fields are known, which is at  ${\cal N}\leq 2$.  The construction of supergravity counterterms, as  initiated in 
\cite{Kallosh:1980fi,Howe:1980th}, based on on-shell classical superspace,  appears to be naive in this respect. Namely, it was more recently shown  in \cite{Chemissany:2012pf}, using ${\cal N}= 2$ supergravity with higher derivatives, that when auxiliary fields are integrated out, the supersymmetry transformations of the physical fields are deformed due to presence of higher derivative terms. This fact makes 
 the predictions of counterterms, based on on-shell classical superspace for ${\cal N}\geq 3$,   inconclusive.

The nonlinear global symmetries of the classical action, in general,  have not been carefully and systematically investigated with the purpose to find out   how they constrain the UV divergences, or even how they constrain the tree level theory.  In supergravity models, in addition to local linear symmetries, there are global nonlinear symmetries, e. g. the $E_{7(7)}$ in ${\cal N}= 8$ supergravity. These symmetries are poorly understood, the lore being that they affect only  the soft limit on amplitudes.
The current conclusion in \cite{Beisert:2010jx} is that starting with 7 loops the candidate counterterms are not forbidden by the soft limit on scalars due to $E_{7(7)}$ symmetry. Meanwhile
the effect of the $E_{7(7)}$ symmetry in the vector sector was also studied in \cite{Kallosh:2011dp,Kallosh:2011qt,Bossard:2011ij,Kallosh:2012yy,Gunaydin:2013pma}. A better understanding of the complete consequences of $E_{7(7)}$ symmetry for ${\cal N}= 8$ supergravity would be desirable.

One of the advantages of the background field method is that the classical symmetries of the action are associated with the 
symmetries of the background field in the context of the effective action. They restrict tree level amplitudes and might, in absence of anomalies, restrict the quantum action. In this paper we will make the first step towards extending the background field method to global nonlinearly realized symmetries. Already at the tree level the situation was not systematically explored in the past, the results in this paper will be restricted to the tree level. A generalization to loop level is outside the scope of this work.

This step will be sufficient to treat models like DBI-VA model and, partially,  ${\cal N}= 8$ supergravity with $E_{7(7)}$ symmetry, with exception of vectors,  using the  general background formalism which will be developed here.

\section{Review of Linear Gauge Symmetries}

In DeWitt's formalism the set of fields $\phi^i$ includes all fields of a given model, summation convention for repeated indices include integrations over the $x$, space-time coordinates. The action $S(\phi)= \int d^4 x {\cal L}(\phi)$ is invariant under a local  symmetry.

The {\it linear gauge symmetry of the action} in the general formalism of DeWitt is a statement that the action is invariant under the transformations
\be
\delta_\xi \phi^i= R^i_\alpha (\phi) \xi^\alpha(x)\, .
\label{lin}\ee
Examples are given in \rf{non-ab} and in \rf{grav}. The gauge symmetry of the action in these notation means that
\be
\delta S (\phi)= {\delta S\over \delta \phi^i} \delta_\xi \phi^i= {\delta S\over \delta \phi^i} R^i_\alpha (\phi) \xi^\alpha =0 \qquad \Rightarrow \qquad {\delta S\over \delta \phi^i} R^i_\alpha=0\, .
\ee
Here the lhs of EOM is defined as follows (in absence of higher derivatives)
\be
{\delta S\over \delta \phi^i} \Rightarrow  \Big({\partial \cL\over \partial \phi^i}- \partial_\mu {\partial \cL\over \partial \partial_\mu \phi^i}\Big )\, .
\ee
Here we use the fact that the parameter of the gauge transformations $\xi^\alpha= \xi^\alpha(x) $ is a function of $x$ and therefore can be dropped from the statement that  $\delta S (\phi)=0$.
The linear symmetry means that
\be
  R^i_\alpha (\phi)= R^i_{\alpha } (0)+ R^i_{\alpha, j}(0) \phi^j
\label{lin}\ee
where the function $R^i_\alpha (\phi)$ has a field independent part $R^i_{\alpha } (0)$ and a linear part $R^i_{\alpha, j}(0) \phi^j$ where $R^i_{\alpha, j} $ is field independent. The local symmetries of the action   form an  algebra:
\be
\{  R^i_{\alpha, j} (\phi)   R^j_\beta (\phi) ] = f_{\alpha \beta}^\gamma  R^i_\gamma (\phi) \, .
\label{algebra}\ee
Here the notation $\{ \cdots ]$ indicates that some elements in the algebra can be  bosonic and some  fermionic.
In case of non-Abelian gauge theory and/or gravity models, the inhomogeneous part $R^i_{\alpha } (0)$ is a differential operator, a
field-independent part of the gauge transformation, 
\be
 R^i_{\alpha } (0) \qquad \Rightarrow  \qquad \delta A_\mu= \partial_\mu \xi(x)\, , \qquad \delta g_{\mu\nu}= \partial_{(\mu} \xi_{\nu)}(x)
\ee
The term linear in fields in   \rf{lin} depends only on the first power of the field.

The action has a quadratic part and an interaction part, 
\be
S= {1\over 2} \phi^i S_{,ij}^0 \phi^j + S_{\rm int} (\phi) = \sum_{n=2} {1\over n!} S_{, i_1\dots i_n} \phi^{i_1} \cdots \phi^{i_n}
\label{full}
\ee
The functions $S_{ ,i_1\dots i_n}$ are functional derivatives of the action over the fields, and therefore they are local functions.

Amplitudes are computed for the on-shell S-matrix and the symmetry acting on a solution of the free field equations has to be used. These are {\it asymptotic symmetries} of the form
\be
 \delta_\xi \phi^i_0= \Big (R^i_\alpha (0) + R^i_{\alpha, j}(0) \phi^j_0\Big ) \xi^\alpha
\label{asymptotic}\ee
Namely, consider the linearised action $S^{(2)}= {1\over 2} \phi^i S_{,ij}^0 \phi^j$, the free field satisfies the free equations of motion
\be
S_{,ij}{}^0 \phi^j_0=0
\label{free}\ee
Even when terms with higher derivatives are added to the classical action, we  expect that such terms start with at least 3 powers of the field, (the terms with 2 powers would break unitarity) and thus, do not affect the form of the asymptotic free equation.
Assuming that $\phi^j_0$ describes a physical state, for example the transverse
 physical vectors and transverse traceless gravitons, we may ignore the field-independent part of the symmetry. This will be dramatically different in case of nonlinear symmetries where $R^i_{\alpha } (0)$ is a constant matrix and $\xi^\alpha $ is a  bosonic or fermionic constant parameter.
 
The important part which controls the properties of the S-matrix for linear symmetries is the second term in \rf{asymptotic}, which for free fields entails the following homogeneous transformation
\be
\tilde \delta_\xi \phi^i_0= R^i_{\alpha, j} \phi^j_0 \xi^\alpha
\label{multiplet}\ee
where $R^i_{\alpha, j}$ is a constant matrix.
For example, in case of a linearized local  supersymmetry, this transformation explains why the $n$-point amplitude with a fixed number of points has information about all members of the susy multiplet, one physical state being transformed into another physical state, starting from helicity $+2$ all the way to helicity $-2$ in ${\cal N}=8$ supergravity. Also in global supersymmetry case, the amplitude with a fixed number of points has information about all members of the susy multiplet
from helicity $+1$ all the way to helicity $-1$ in ${\cal N}=4$ YM theory.  In this case the algebra \rf{algebra} is the linear global supersymmetry algebra.

With regard to restrictions on the quantum action, the linear gauge symmetries are known to constrain UV divergences which is a rather well known procedure of implementing Ward identities on higher derivative counterterms. For example, the form of the UV loop divergences in super-YM theory and gravity was deduced using the background field method \cite{DeWitt:1967ub} and confirmed by actual computation, \cite{'tHooft:1974bx,Kallosh:1974yh,Grisaru:1975ei,Goroff:1985th,vandeVen:1991gw}.

Note that in contrast to nonlinear symmetries, the linear symmetries do not involve a constant term in the transformation rules and also have a property
\be
R^i_{\alpha, jk} (\phi)=0
\ee
i.e. there are no quadratic or higher powers of the fields in the transformation rules.

\section{General  nonlinear global symmetry models}

Here we will develop the {\it background field method  for global nonlinear symmetry, which can be bosonic or fermionic}. 
In global symmetry 
\be
\delta\phi^i= {\cal R}^i_\alpha (\phi) \xi^\alpha
\label{NLS}\ee 
the parameters $\xi^\alpha$ are space-time independent, bosons for bosonic symmetries and fermions for fermionic ones. The nonlinearity means that in addition to a field independent and first order in field as in \rf{lin}, the transformation has higher order in field terms, at least quadratic
\be
 {\cal R}^i_\alpha (\phi)_{\rm nonlin}= {\cal R}^i_\alpha (0) + {\cal R}^i_{, j \alpha}(0)  \phi^j  \xi^\alpha + {\cal R}^i_{,jk\alpha} (0) \phi^j \phi^k  +\dots 
\ee
Here the field independent part of the transformation is a constant matrix
\be
 {\cal R}^i_\alpha (0) \xi^\alpha = {\rm const}
\ee

When the action has a symmetry under some global transformation of fields, it means that the variation of the Lagrangian under global symmetries is a total derivative and can be presented in the form
\be
\delta \cL={\delta S\over \delta \phi^i} \cR^i_\alpha (\phi) \xi^\alpha+ \partial_\mu \cJ^{\mu \, N}= \partial_\mu {\cal J}^\mu
\label{sym}\ee
where the Noether current is defined
\be
\cJ^{\mu \, N}\equiv  {\partial \cL\over \partial \partial_\mu \phi^i} \cR^i_\alpha (\phi) \xi^\alpha
\ee
Thus, when EOM following from the given action are satisfied, the current conservation follows
\be
{\delta S\over \delta \phi^i}=0 \qquad \Rightarrow \qquad \partial_\mu \cJ^{\mu \, N}- \partial_\mu {\cal J}^\mu=0  
\label{current}\ee
Under appropriate boundary conditions one finds that 
\be
\delta S= \int \delta \cL= \int \partial_\mu {\cal J}^\mu =0
\label{symmetry}\ee

Note that  here ${\cal R}^i_\alpha (\phi)$ are some local nonlinear functions of the field $\phi$. The nonlinear symmetry is associated with spontaneous symmetry breaking, which means that the transformation law has a constant field-independent part
$ {\cal R}^i_\alpha (0) \xi^\alpha $.
The transformations form some open algebra
\be
\{ {\cal R}^i_{\alpha, j} (\phi)  {\cal R}^j_\beta (\phi) ] = f_{\alpha \beta}^\gamma {\cal R}^i_\gamma (\phi) +\eta^{[ij]}_{\alpha \beta} {\delta S\over \delta \phi^j}
\label{al}\ee
The first term here defined by the structure constants is the usual one, the second term in the case of nonlinear symmetries, proportional to field equations, often shows up in nonlinear supersymmetries, see for example \cite{Kuzenko:2011tj} for Goldstino action.

\subsection{S-matrix}

The path integral describing the  S-matrix is given by the following expression
\be
e^{i\Gamma(\phi_0)} = N \int d\Phi e^{i \Big (S(\Phi) + \phi_0^i S_{,ij}{}^0 \Phi^j\Big )}
\ee
Here $\Gamma (\phi_0)$ describes a functional for all connected S-matrix amplitudes.
\be
\Gamma(\phi_0)= \sum_{n=4} ^\infty {1\over n!} \, A_{i_1 \dots i_n} \phi^{i_1}_0 \dots \phi^{i_n}_0
\label{Gamma}\ee
For example, the 4-point S-matrix amplitude is $A_{i_1 i_2 i_3 i_4}$, the 6-point S-matrix amplitude is $A_{i_1 i_2 i_3 i_4 i_5 i_6}$ etc.
  The field $\phi_0^i$ is a solution of the free equations of motion \rf{free}.
In case that also gauge symmetries are present, the corresponding gauge-fixing procedure is in order. We will focus here on a case without local symmetries\footnote{Local symmetries in models of the Born-Infeld type, when the action depends only on a gauge-invariant field strength $F_{\mu\nu}$, can also be easily described in this formalism.}
 but with nonlinear global symmetries. The background field $\phi$ satisfies classical field equations
\be
{\delta S\over \delta\phi^i}=\phi_0^j \vec S_{,ji}{}^0 \, , 
\label{EOM}\ee
see,  \cite{DeWitt:1967ub,Honerkamp:1971sh}, \cite{Kallosh:1974yh}. The right hand side  of the equation of motion ${\delta S\over \delta\phi^i}$ is  viewed as an operator: if this equation is applied to an object $X^i= G^{ik}_0 Y_k$ which has a pole,  we find $\phi_0^i \vec S_{,ij}{}^0G^{jk}_0 Y_k =-\phi_0^i Y_i $.
Here $G^{ij}_0$ is a free propagator, inverse to a differential operator $ S_{,ij}{}^0$, 
\be
S_{,ij}{}^0 G^{jk}_0= - \delta^i_k
\ee
As discussed above, even with higher derivatives terms added, $ S_{,ij}{}^0$ should not be modified, to preserve unitarity.
The background field $\phi(\phi_0)$, which solves eq. \rf{EOM} is defined by iteration
\be
 \phi^i  = \phi^i_0 + G^{ij}_0 \sum_{n=2} {1\over n!} S_{, j i_1\dots i_n}\phi^{i_1} \dots \phi^{i_n}
\label{sol}\ee
 We can also write \rf{sol} as
\be
 \phi^i  = \phi^i_0 + G^{ij}_0 {\delta S_{\rm int}(\phi)\over \delta \phi^j}
\label{useful}\ee
Note that the background field defined above  as a function of the free field does depend on the additional terms in the action with higher derivatives, which affect $S_{\rm int}$.
The iterated solution of \rf{sol} for the background field $\phi$, which is functional of a free field $\phi_0$, yields all tree functions
\be
\phi^i[\phi_0]\equiv \phi_0^i+ G^{ij}_0 \sum_{n=2}^{\infty} {1\over n!} t_{ji_1\dots i_n} \phi^{i_1}_0 \dots \phi^{i_n}_0\, , 
\label{tree1}\ee
see,  for example, eq. (14.2) in \cite{DeWitt:1967ub}. Here the tree functions $t_{ji_1\dots i_n}$ are given by a set of propagators, describing the tree, they are highly non-local. 

The path integral defining all  tree diagrams  is given by the following expression, \cite{DeWitt:1967ub}, \cite{Kallosh:1974yh}
\be
e^{i\Gamma(\phi_0)} =  e^{i\Big (S(\phi)- S_i \phi^i \Big )} \Big |_{\phi=\phi[\phi_0]}
\label{Smatrix}\ee
where the background field $\phi$ is given as a functional of the free field $\phi[\phi_0]$ as defined in \rf{tree1}. This is just a saddle point expansion of the path integral where all loop corrections are neglected.
All connected tree diagrams are given by the  functional presented in eq. \rf{Gamma}.

As we have shown, in terms of the background field $\phi=\phi[\phi_0]$ the sum of all n-point amplitudes can be also given in the form

\be\boxed {
\Gamma(\phi_0)= \sum_{n=4} ^\infty {1\over n!} \, A_{i_1 \dots i_n} \phi^{i_1}_0 \dots \phi^{i_n}_0= \big (S(\phi)- S_i \phi^i \Big ) \Big |_{\phi=\phi[\phi_0]}}
\label{GammaNL}\ee
 As a function of the background field $\phi$ the $S$-matrix functional, which is a sum of all amplitudes, is invariant under a nonlinear symmetry
\rf{NLS}.

Thus, the sum of all n-point amplitudes $A_{i_1 \dots i_n} $ (local and non-local parts) has to produce an expression above in \rf{GammaNL} which has a complete information on the nonlinear symmetry of the theory and 
its algebra \rf{al}. Moreover, as a function of the background field, it is a local expression.

This means that various soft theorems  inferred from  explicitly constructed amplitudes  $A_{i_1 \dots i_n} $
 originate from a universal eq. \rf{GammaNL}. In other words: a relation between amplitudes with different number of points, which was often times obtained in the literature using explicit expressions for $A_{i_1 \dots i_n} $ in different models, actually follows from the fact that all these $n$-point amplitudes have to sum up into an expression in the rhs of eq. \rf{GammaNL}, which is invariant under nonlinear symmetry transformations. These, in turn, inevitably, relate amplitudes with different number of points, since the constant term in $\delta\phi$ decreases this number, the linear term preserves it, a quadratic term increases the number of points by one, etc. Moreover, when $\phi$ is replaced by the full tree formula \rf{tree1} in terms of the free field, the relations between different point amplitudes become even more involved. 
 
  It has been observed in \cite{ArkaniHamed:2008gz,Brodel:2009hu} that for  ${\cal N}= 8$ supergravity the double-soft limit of two scalars probes the coset space structure of the  spontaneously broken $E_{7(7)}$  symmetries producing an unbroken $SU(8)$ symmetry. Not accidentally, therefore, the soft limit theorems were associated with the `footprints' of the nonlinear symmetries in \cite{Kallosh:2008rr}. An analogous property of the double-soft limits in VA theory were observed in amplitudes computed in \cite{Chen:2014xoa}. Here we would like  to find out if this feature of the double-soft limit follows from the nonlinear symmetry, in general.

\subsection{Nonlinear symmetry of the background field versus linear symmetry of the free field}
The nonlinear global symmetry of the action $S(\phi)$ for the background field $\phi$ solving classical equations of motion \rf{EOM}, or more complicated ones in case of higher derivative terms added to the classical action,  is defined by the expression in \rf{NLS}. We are interested in asymptotic symmetry of the free field $\phi_0$ satisfying eq. \rf{free}. In the asymptotic limit action only the quadratic term survives:
\be
S^{(2)}= {1\over 2} \phi^i_{\rm as} S_{,ij}^0 \phi^j _{\rm as}
\label{2}
\ee
A zero mode of the eq. 
$
S_{,ij}{}^0 \delta \phi^i_0  =0
$
is therefore the symmetry of the free field. It is also an asymptotic limit of the full symmetry of the action, thus
the transformation of the free field include the terms of the zero and first power in field, namely
\be
\delta_\xi \phi^i_0  =  R ^i_\alpha(\phi_0) \xi^\alpha \equiv R ^i_\alpha (0) \xi^\alpha + R ^i_{ \alpha, j}(0) \phi_0^j \xi^\alpha
\label{linear} \ee 
The new situation here, comparative to linear symmetries we discussed in sec. 2. is the case that the free field is shifted by a boson, for a bosonic field, and by a fermion, for a fermionic field.
\be
\delta_\xi^{\rm shift} \phi^i_0= R ^i_\alpha (0) \xi^\alpha
\label{shift1}\ee
by the first term in eq. \rf{linear}.  In case of linear symmetries in gauge theories, YM and gravity, the field independent term correspond to a free gauge transformation and is trivial for transverse vectors and gravitons. For example, $\delta A_\mu= \partial_\mu \Lambda$ in general, but if the free field is transverse, $\delta A_\mu^{\rm transverse}=0$ and we find that the field independent part of the transformation vanishes for the physics state described by the solution of the free field equation, 
$
\delta \phi_0^i = R^i_\alpha (0) \xi^\alpha =0
$

Now, in case of global nonlinear symmetries,  we consider an opposite case, the effect of the field-independent shift of the free field is not trivial, i. e.
\be
\delta^{\rm shift}_\xi   \phi_0^i = R^i_\alpha (0) \xi^\alpha \neq 0
\label{shift1}\ee
This transformation is the essence of spontaneous symmetry breaking, when there is a field-independent term in the transformation rules.
Now the symmetry under \rf{linear}  is a symmetry of the S-matrix functional $\Gamma(\phi_0)$ in \rf{Smatrix} under asymptotic symmetry of the free field, 
\be
\delta_\xi \Gamma(\phi_0)=\sum_{n=4} ^\infty {1\over (n-1)!}A_{i_1 \dots i_n} \phi^{i_1}_0 \dots \phi^{i_{(n-1)}}_0 R^{i_n} _{\alpha}(\phi_0) \xi^\alpha =0
\label{result}\ee
Thus, the asymptotic linear symmetry means that
\be 
\sum_{n=4} ^\infty {1\over (n-1)!}A_{i_1 \dots i_n} \phi^{i_1}_0 \dots \phi^{i_{(n-1)}}_0 \Big ( R^{i_n} _{\alpha}(0) +R ^{i_n}_{ \alpha, j}(0) \phi_0^j\Big )  \xi^\alpha =0
\label{general}\ee
Since the first term decreases the number of fields by one, and the second term preserves it, we find a relation between the $n$-point amplitude and the $(n-1)$-point amplitude:
\be 
\Big [ {1\over (n-1)}A_{i_1 \dots i_n} \phi^{i_1}_0 \dots \phi^{i_{(n-1)}}_0  R^{i_n} _{\alpha}(0)    +A_{i_1 \dots i_{n-1}} \phi^{i_1}_0 \dots \phi^{i_{(n-2)}}_0 \phi_0^{j} R ^{i_{n-1}}_{ \alpha, j}(0)  \Big ]\xi^\alpha=0
\label{n and n-1}\ee
Here the symmetry transformation of the asymptotic free field \rf{linear}, codified by the expressions $ R^{i} _{\alpha}(0) $ and $R ^i_{ \alpha, j}(0)$ is the only ingredient in the relations between the $n$-point and $n-1$ point amplitudes.

\section{Simple case of `even only' amplitudes}
Consider the models where only even amplitudes are non-vanishing, $n=2k$ where $k$ is an integer. For example all-fermion amplitudes, or other models where  odd number of points in the amplitude is not possible \footnote{In non-Abelian gauge theories and gravity and supergravity it is possible to add gauge fields and gravitons, therefore, there are amplitudes with even and odd number of points. In such case,  the restriction \rf{n and n-1} should be applied, in general.}. This class of models was recently studied in \cite{He:2016vfi}, \cite{Cachazo:2016njl}. They include $U(N)$ nonlinear sigma model (NLSM),  DBI-VA model and some other models.

The asymptotic linear symmetry of the amplitude functional in case of even amplitudes, $n=2k$, is reduced to the following two constraints. In terms of connected tree diagrams in \rf{Gamma} the statement in 
\rf{result}-\rf{n and n-1} means that separately we find restrictions on $n=2k$ amplitude. The first from the shift part, the second from the homogeneous part

\be 
A_{i_1 \dots i_n} \phi^{i_1}_0 \dots \phi^{i_{(n-1)}}_0  R^{i_n} _{\alpha}(0) \xi^\alpha   =0
\label{Adler1}\ee

\be
A_{i_1 \dots i_{n}} \phi^{i_1}_0 \dots \phi^{i_{(n-1)}}_0 R ^{i_{n}}_{ \alpha, j}(0) \phi_0^{j} \xi^\alpha=0
\label{homogen}\ee

\subsection{Adler's zero, for bosons and fermions}
 The symmetry of the S-matrix under the shift $\delta^{\rm shift}_\xi   \phi_0^i= R^{i} _{\alpha}(0) \xi^\alpha$ can be considered independently of the homogeneous linear transformation $\tilde \delta_\xi  \phi^i _0= R ^{i}_{ \alpha, j}(0) \phi_0^j \xi^\alpha$ in models with only even point amplitudes. In such case
\be
\delta_\xi ^{\rm shift}\Gamma(\phi_0)=0
\label{resultShint}\ee
under the shift is the statement about {\it Adler's zero}  \cite{Adler:1964um}, \cite{Weinberg:1966kf} for any bosonic or fermionic symmetry which has a constant field independent term in the symmetry transformation.

The linear symmetry of the amplitude functional in case of even amplitudes, $n=2k$ where $k$ is integer, is reduced to the following two constraints.
In terms of connected tree diagrams in \rf{Gamma} the statement in 
\rf{resultShint} means that as shown in eq. \rf{Adler1},
 when one of the free fields $\phi_0^{i_n}(x) $ in the effective action for the $n$-point amplitude is replaced by a constant $R ^{i_n}_\alpha (0) \xi^\alpha$, the amplitude vanishes. This means in the momentum space that in a single soft limit every $n$-point amplitude vanishes.
 
 Thus, for models with only even amplitudes, like the DBI-VA model, the VA model, the $U(N)$ nonlinear sigma model etc, we just proved that they have Adler's zero in a single soft limit for all states whose symmetry transformation contains a constant term. In DBI-VA model there are such constant shifts in VA supersymmetry of fermions, there is a shift by a constant on scalars and on vectors, since the action depends only on derivatives of scalars and vectors, see Appendix for details. Thus we conclude that in DBI-VA model the amplitudes vanish in the single soft limit for any physical state.

The  simplicity and universality of this derivation is surprising. 
The original Volkov-Akulov derivation of the Adler's zero in \cite{Akulov:1974xz} involves a significant amount of work, but the  tools did not involve the background field method. The single and double soft limit of VA theory were also studied in \cite{deWit:1975xci}, where it was shown, for the first time, that the double soft limit is defined by the symmetry commutators.

It is also interesting that in  the DBI-VA model studied recently  in  \cite{He:2016vfi}, \cite{Cachazo:2016njl}  the single soft limit for any choice of external states vanishes. To see this one can
use the first columns of Table I and  II in \cite{Cachazo:2015ksa}. In this way the AdlerÕs zero for a soft scalar in various EFTÕs can be read off by studying soft behavior of their CHY formulas. By performing the same analysis in 4d, one can we see that $A_{n=2k} \to O(\tau) $ as the soft momentum of a scalar, fermion or photon becomes soft with $\tau\to 0$. One can also see this explicitly from eq. (5.10) in \cite{Cachazo:2016njl}. The important fact we would like to stress here is that the proof of Adler's zero in amplitudes is based on the knowledge of amplitudes and is inferred from their explicit expressions in  \cite{He:2016vfi}, \cite{Cachazo:2016njl}.

{\it In our proof we never use the explicit knowledge of amplitudes, but we use the symmetry of the action in the framework of the background field method.   The crucial feature for the universal proof of the vanishing of the soft limit for bosons and fermions is due to the presence of a constant term in the symmetry transformation. In general, we have derived a relation between the soft limit of the $n$-point  amplitude and the $(n-1)$-point amplitude, see \rf{general}, \rf{n and n-1}. But for models with only even  amplitudes  a single soft limit of the $2n$-point amplitude always vanishes, universally, when symmetry involves a constant shift.}

\subsection{A multi-component physical state}
The constraint shown in \rf{homogen} has a simple interpretation: When the physical state, described by the free field $\phi^i_0$ is a multi-component state, the symmetry in \rf{homogen} relates the $n$ point amplitudes with states in the multi-component field. For example, in case of linear supersymmetry it relates $n$-point amplitudes in the same supermultiplet. A well know example, used also recently  in  \cite{He:2016vfi} is based on amplitudes in application to nonlinear symmetries in DBI-VA model. They depend on a manifest ${\cal N}= 4$ supersymmetry of the physical state given by an on-shell  superfield for photons, photinos, and scalars
\be
\Phi^{DBI-VA}(\eta)= \gamma^+ + \eta^A \psi_A  +{1\over 2!} \eta^A \eta^B S_{AB} + {1\over 3!} \eta^A \eta^B  \eta^C \epsilon _{ABCD}\bar \psi^D + \eta^1\eta^2 \eta^3 \eta^4 \gamma^- 
\ee
The reason why using this multiplet represents an example of eq.  \rf{homogen} is that any $2n$-point amplitude describes all states of the multiplet, the relation between them corresponds to the symmetry which in general is given in eq.  \rf{homogen}. This is still a manifestation of the linear part of the symmetry, it does not mix the amplitudes with different number of points.

\section{Nonlinear symmetry and relation between amplitudes with different number of points}

So far we have only identified the effect of the constant term in the nonlinear symmetry transformation, $\delta ^{\rm shift} \phi^i_0$, \rf{shift1}  as well as the linear field-dependent transform $\tilde \delta$ in \rf{multiplet}. It is particularly simple in models with `even only' amplitudes where it
 leads to a universal and simple proof of the vanishing single soft limit in all tree level S-matrix amplitudes of a given number of points, as well as a symmetry under a group transformation. For a fixed number of points the linear symmetry of the form \rf{multiplet} relates amplitudes of a given multiplet to each other, with the same number of points.

The main effect of nonlinearity, in addition to vanishing single soft limit in all tree level S-matrix amplitudes follows from a fact that the on-shell S-matrix $\Gamma(\phi_0)$ can be also presented as a function of the background field $\phi$, as shown in eq. \rf{GammaNL}. The fact that under transformation of the background field \rf{NLS} the action is invariant, when field equations are satisfied, see \rf{current}, \rf{symmetry}
\be
\delta_\xi S(\phi) =0\, , \qquad \delta_\xi \phi^i = {\cal R}^i_\alpha (\phi) \xi^\alpha= {\cal R}^i_\alpha (0) \xi^\alpha + {\cal R}^i_{\alpha, j  }(0)  \phi^j  \xi^\alpha + {\cal R}^i_{\alpha ,jk } (0) \phi^j \phi^k \xi^\alpha +\dots 
\label{nonlin}\ee
means that terms in the action with different powers of $\phi$ are related, due to the nonlinearity of the symmetry \footnote{It was stressed in \cite{DeWitt:1967ub}, \cite{Kallosh:1974yh,Grisaru:1975ei} that the background method developed in these papers is valid for ${\cal R}^i_{\alpha ,jk }(\phi)=0$, so it applies only to  linear symmetries.}.
 When the dependence on $\phi$ is replaced via eq. \rf{tree1} the relation between different powers of $\phi_0$ in the expansion of 
\be
 \sum_{n=4} ^\infty {1\over n!} \, A_{i_1 \dots i_n} \phi^{i_1}_0 \dots \phi^{i_n}_0= S\Big (\phi(\phi_0)\Big ) -\phi_0^i \vec S_{,ij}{}^0 \phi^j (\phi_0)
\label{Identity}\ee
may be deduced, in principle. 

The relation between the  single soft limit and double soft limit probing the coset space structure of models with spontaneously broken symmetry and the background field method proposed here needs to be studied in various models. It is important to stress, however, that the constraint on amplitudes presented in \rf{GammaNL},  \rf{Identity} was derived from the path integral saddle point expansion and the rhs of the constraint \rf{Identity} has the complete information about the nonlinear (super) symmetries of the model. Therefore it is not surprising that the algebra of the symmetries shows up universally in the double-soft limit.

It might be useful to remind the following identity. The symmetry of the action in the form of eq. \rf{sym} can be given in the form
\be
S_{, i } {\cal R}^i_\alpha \xi^\alpha = \partial_\mu ({\cal J }- {\cal J}^N)^\mu
\ee
We can now perform the variation of this identity over the fields taking into account that there is no variation from the rhs of this equation.
We find that
\be
(S_{, i_1 i_2} {\cal R}^{i_1}_\alpha + S_{, i_1 } {\cal R}^{i_1}_{\alpha , i_2 })\xi^\alpha =0
\label{major}\ee

\subsection{Double-soft limit}
Consider a nonlinear symmetry transformation of the identity \rf{major} with the parameter $\xi'$. We find that
\be(S_{, j i_1 i_2} {\cal R}^{i_1}_\alpha  {\cal R}^{i_2}_\beta+S_{,  i_1 i_2} {\cal R}^{i_1}_{ \alpha, j } {\cal R}^{i_2}_\beta+ S_{, j i_1 } {\cal R}^{i_1}_{\alpha , i_2 } {\cal R}^{i_2}_\beta+S_{, i_1 } {\cal R}^{i_1}_{ \alpha , j i_2 } {\cal R}^{i_2}_\beta)\xi^\alpha  \xi^{'\beta} =0
\ee
We can also present this expression separating the algebra term with the structure constant and terms depending on  $S_{,i}$ which have to be taken into account carefully, before dismissing:
\be
\Big (S_{, j i_1 i_2} {\cal R}^{i_1}_\alpha  {\cal R}^{i_2}_\beta+ S_{, j i_1 } {\cal R}^{i_1}_{ \gamma} f^\gamma_{\alpha \beta} + S_{,i_1} ({\cal R}^{i_1}_{\alpha , j i_2} {\cal R}^{i_2}_{ \beta}- {\cal R}^{i_1} _{\beta , i_2} {\cal R}^{i_2}_{ \alpha , j}\Big )\xi^\alpha  \xi^{'\beta} =0
\label{doublesoft}\ee
This identity might be useful for understanding the universality of the double-soft limit. In the approximation when we keep in the first term in \rf{doublesoft} only the field-independent constant term, it is equivalent to taking a double-soft limit. The remaining   terms in this identity define this double-soft limit. For example  the second term depends universally on the algebra of two transformations with spontaneously broken symmetry which has such constant terms.

\subsection{Triple- and multiple-soft limit}
Now we perform the symmetry variation of the identity above, namely, consider a nonlinear symmetry transformation with the independent parameter $\xi^{''}$. The result is
\be
\Big (S_{, j i_1 i_2 i_3} {\cal R}^{i_1}_\alpha  {\cal R}^{i_2}_\beta {\cal R}^{i_3}_\gamma+ \cdots \Big )\xi^\alpha  \xi^{'\beta} \xi^{''\gamma}=0
\label{triple}\ee
All terms in $\cdots$ can be written down by performing a variation of each term in \rf{doublesoft} over $\phi^{i_3}$ and by multiplying the result by the symmetry variation ${\cal R}^{i_3}_\gamma \xi^{''\gamma}$. The first term has a part where each ${\cal R}^{i_1}_\alpha$ and ${\cal R}^{i_2}_\beta$ and ${\cal R}^{i_3}_\gamma$ are constant matrices, reflecting the constant term in the symmetry, like $\delta_\xi^{\rm shift} \phi^i_0= R ^i_\alpha (0) \xi^\alpha$ in eq. \rf{shift1}. This terms is associated with the triple-soft limit, whereas all other define its value according to identity \rf{triple}.

For the multiple-soft limit one finds an analogous type constraint, after performing multiple symmetry transformation so that
\be
\Big (S_{, j i_1 \dots i_n} {\cal R}^{i_1}_{\alpha_1}    \dots  {\cal R}^{i_n}_{\alpha_n} + \cdots \Big )\xi^{\alpha_1}   \dots  \xi^{\alpha_n} =0
\label{multi1}\ee
Here the first term where we keep only $\delta_\xi^{\rm shift} \phi^i_0= R ^i_\alpha (0) \xi^\alpha$ is the multiple-soft limit, the remaining terms define its value according  to identity \rf{multi1}.

\

\

 To summarize, one can view the identity  \rf{Identity} for the sum of the amplitudes of all number of points together with nonlinearity of the symmetry transformation of the background field shown in \rf{nonlin} which restricts the rhs of the identity \rf{Identity}, as a generalization of the standard background method used in \cite{DeWitt:1967ub}, \cite{Kallosh:1974yh,Grisaru:1975ei} in case of linear symmetries. 
 
 The new feature here is the presence of a constant field independent term  in \rf{nonlin}, which in the most general case defined a single soft limit for an $n$-point amplitude via an $n-1$ amplitude as shown in \rf{n and n-1}. In particular, when there are only even amplitudes, it means that in the single soft limit all amplitudes vanish. Another new feature is that the identity   \rf{Identity} encodes a complete information about a nonlinear symmetry and its algebra, and  requires some relation between amplitudes of various number of points.
 
 These relations can be seen, for example, as the soft limit on the amplitudes, related to the soft limit of the background field described above.

\section{Short comments on  recent relevant papers} 

The studies of amplitudes in VA theory were initiated in \cite{Chen:2014xoa}.
The 4-point and 6-point amplitudes of VA  model have been computed recently in \cite{Luo:2015tat} and confirmed in \cite{He:2016vfi,Cachazo:2016njl}. Moreover, a complete all-tree amplitude expression was derived in  \cite{He:2016vfi,Cachazo:2016njl}. It was also stressed that the case of maximaly supersymmetric DBI-VA theory leads to a nice and elegant S-matrix formula, comparative with other models with nonlinear symmetries where also the new S-matrix expression were derived.

Using the new amplitudes it was shown in  \cite{He:2016vfi,Cachazo:2016njl} that all amplitudes in an one-particle soft limit vanish, whereas in the double soft limit  $n$-point amplitude is related to the $n-2$ point amplitude via an expected relation \cite{ArkaniHamed:2008gz,Brodel:2009hu} associated with the algebra of two nonlinear supersymmetry transformation.

The fermions in  DBI-VA are in the $SU(4)$ $R$-symmetry group. A choice of the fermion amplitudes was made as follows: the fermions $\psi$ were taken with the R-charge (123) and the fermion $\bar \psi $ with a charge (4). This results in fermion amplitudes of the same type we would get in ${\cal N}=1$ pure VA theory. Namely, in notations of  \cite{Cachazo:2016njl}
\be
A_4 ^{DBI-VA} (1^\psi, 2^{\bar\psi}, 3^\psi, 4^{\bar\psi})= - s_{13}  \langle 24\rangle [13]
\label{A4}\ee
This expression is unique, by helicity and dimension. The 6-point amplitude defined as $A_6 ^{DBI-VA} (1^\psi, 2^{\bar\psi}, 3^\psi, 4^{\bar\psi}, 5^\psi, 6^{\bar\psi})$ is also given by a rather compact formula in which every term out of 9 has a factor like $A_4 ^{DBI-VA}$ times some other functions. Moreover, the property of the single and double soft limits follow from the 6-point amplitude by a direct observation. Therefore the soft limits look like footprints of the corresponding symmetries. This idea that soft limits are footprints of nonlinear symmetries was proposed and developed with regard to $E_{7(7)}$ symmetry in ${\cal N}=8$ supergravity in \cite{Kallosh:2008rr}. Here we have done more, we have shown how the nonlinear symmetry affects the amplitudes directly, via eq. \rf{GammaNL}, not just via soft limits.

 The double-soft limit in DBI-VA model  relates amplitudes with two soft particles of spin  $s=0,1/2, 1$ to   amplitudes without these 2 particles. The small  parameter $t \rightarrow 0$ is introduced as follows
$\bar \lambda (p) \rightarrow  t \bar \lambda (p)$, $ \lambda (q) \rightarrow  t \lambda (q)$
\be
{\cal M}^{(s)}_{n+2} = t^{1+2s} \sum _{a=1}^{n} {(k_a \cdot (q-p))^{2-2s}\over 2k_a\cdot (p+q)} [p | a | q  \rangle^{2s} {\cal M}^s_n + {\cal O} (t^{2+2s})
\label{double1}\ee
Here we use the results in the form given in \cite{He:2016vfi}. For $s=1/2$, for soft fermions,  one can see that if either $q$ or $p$ actually vanish so that $ {(k_a \cdot (q-p))\over 2k_a\cdot (p+q)} =\pm 1$, the amplitude vanishes as the single soft limit theorem is predicting.
Also in $s=1/2$ case 
\be
t^2 [p | a | q  \rangle = t \bar \lambda_{\dot \alpha}  (p)   k_a^{\alpha \dot \alpha} t \lambda_\alpha (q)
\ee
In the limit of vanishing $t$ the spinors can be replaced by a momentum independent constant spinors
\be
t^2 [p | a | q  \rangle_{t\rightarrow 0} =  \bar \epsilon_{\dot \alpha}^1     k_a^{\alpha \dot \alpha} \epsilon_\alpha^2 
\ee
The nonlinear supersymmetry \rf{susy} has the following algebra

\be
\delta_{[ \epsilon_2 } \delta_{\epsilon_1 ]} \lambda^\alpha = - i (\epsilon_2 \sigma^\rho \bar \epsilon_1 - \epsilon_1 \sigma^\rho \bar \epsilon_2) \partial _\rho \lambda^\alpha = \xi^\rho \partial _\rho \lambda^\alpha
\ee
corresponding to 
\be
\{ Q_\alpha , \bar Q_{\dot \alpha} \} = P_{\alpha \dot \alpha}
\ee
Thus the non-vanishing of the double-soft limit in \rf{double1} is due to a non-commutative nature of the spontaneously broken symmetry generators, as expected.

\section{Nonlinear symmetry and quantum theory/higher derivative terms}
When quantum corrections in models with nonlinear symmetry are computed, and UV divergences are identified, one can think of adding the relevant counterterms 
to the original action, with the purpose to absorb these infinities. A special property of the nonlinear symmetries is that adding new terms to the action modifies the form of the symmetry. This is different from  gauge models or gravity where the symmetry of the classical action with  higher derivative terms  is the same as in classical theory, see for example eqs. \rf{non-ab} and \rf{grav}.
There are some interesting examples studied in the past which we will shortly describe here.

\subsection{  Born-Infeld with Higher Derivatives}
It was proposed in \cite{Bossard:2011ij} that a consistent procedure for constructing higher derivative terms with duality symmetry on vectors, including the $E_{7(7)}$ symmetry, requires to start with the so-called initial source of deformation, which has to be duality invariant and must depend  on duality doublet with independent  upper and lower components. 

In case of $U(1)$ duality symmetry the corresponding  procedure to produce BI models with higher derivatives was presented in  \cite{Carrasco:2011jv}.
The procedure preserves the $U(1)$ duality of the classical action, by deforming the symmetry simultaneously with deforming the action. In \cite{Chemissany:2011yv} this procedure was explicitly realized, a recursive algorithm was given how to find all higher-order terms in the action of the form $\lambda^n \partial ^{4n} F^{2n+2}$, which are necessary for the $U(1)$ duality current conservation. In these models the duality invariant source of deformation had 4 derivatives and 4 fields.  It was argued in \cite{Chemissany:2011yv} that the class of models with $ \partial ^{2m} F^{2n}$ can also be presented with a complete deformation, so that the $U(1)$ duality is satisfied, starting with more general sources of deformation. For example, one can start with the open string 4-point amplitude with higher derivatives. It was established in \cite{Chemissany:2006qd,deRoo:2003xv} that electromagnetic selfduality is satisfied to all orders in $\alpha'$ for the four-point function sector of the four dimensional open string effective action. Their argument is valid for any number of derivatives in the 4-point function but  up to higher order corrections related to $n$-point amplitudes with $n>4$. Maybe one can try to make progress here  by producing an explicit $U(1)$ duality invariant action with higher derivatives, starting with open string Veneziano-type 4-point amplitude. 

To summarize, the $U(1)$ duality models are not unique, depend on the choice of the BI-type action and the choice of initial source of deformation, and, maybe, it is possible to further clarify some important models of this kind, including the open string theory, with infinite number of derivatives.

\subsection{Duality groups of type E7 and higher derivatives }

Few years ago there was a proposal in \cite{Kallosh:2011dp,Kallosh:2011qt} to sort out  the action of $E_{7(7)}$ symmetry on vectors,  by taking into account that the classical Noether construction \cite{Noether:1918zz} has to be replaced by a NGZ formalism  \cite{Gaillard:1981rj}.
For scalars, in the case of $E_{7(7)}$, the symmetry is described as proposed in sec. 3 of this paper, it is a Noether-type symmetry  \cite{Noether:1918zz}. The conclusion in \cite{Beisert:2010jx}, that starting with 7 loops the candidate counterterms are not forbidden, was based on the scalar soft limit due to $E_{7(7)}$ symmetry. Meanwhile, 
the action of $E_{7(7)}$ on vectors, a duality symmetry, is not described by Noether construction. The  $E_{7(7)}$ symmetry does not act on the field $A_\mu$ (an integration variable in the path integral), but it acts on a duality doublet constructed from the fields strength and a dual to a functional derivative of the action over the field strength, $\Big (F_{\mu\nu}, G^{\mu\nu}\Big)$, where $\tilde G^{\mu\nu} =2 {\delta S\over \delta  F_{\mu\nu}}$. The corresponding identities generalizing the ones by Noether were derived in  \cite{Gaillard:1981rj}. These NGZ identities are complicated, in general,  but they are simplified when $E_{7(7)}$ duality symmetry is consistently truncated to $U(1)$ duality symmetry.

The effect of the $E_{7(7)}$ symmetry in the vector sector was  studied in \cite{Kallosh:2011dp,Kallosh:2011qt,Bossard:2011ij,Kallosh:2012yy,Gunaydin:2013pma}.  In particular, it was observed in \cite{Kallosh:2012yy} that the conclusion concerning $E_{7(7)}$ symmetry in ${\cal N}=8$ supergravity supporting the UV finiteness, in absence of anomalies,  is valid for ${\cal N}\geq 4$ supergravities where duality group is of the type E7 \footnote{It was  explained in \cite{Ferrara:2011dz} that all extended supergravities with ${\cal N}\geq 4$ have duality groups of the type E7. The essence of groups of type E7 is that they have unique quartic invariants for symplectic representations and there is no quadratic invariant for a single duality doublet.}.

After the study in \cite{Kallosh:2012yy} it was discovered in \cite{Bern:2013uka} that the 4-loop UV divergence of ${\cal N}=4$ supergravity might be related to $U(1)$ anomaly, which is a subgroup of the $SL(2, R)$ part of duality \cite{Carrasco:2013ypa}.  It was also discovered that the 4-loop UV divergence of ${\cal N}=5$ supergravity is cancelled, \cite{Bern:2014sna}. Both of this computational facts appear to be in agreement with the predictions made in
\cite{Kallosh:2012yy} based on an analysis of the duality current conservation. In ${\cal N}=4$ supergravity the duality group is $SU(4)\times  SL(2,R)$, which belongs to E7 type. The anomaly of  $SL(2,R)$ makes the formal prediction in \cite{Kallosh:2012yy} of UV finiteness invalid. Note that $SU(4)$ without $ SL(2,R)$ is not of E7 type.

In ${\cal N}=5$ the 10 vector field strengths and their duals form a three-fold antisymmetric irreducible representation  {\bf 20} of the U-duality  $SU(1,5)$. There is a  unique algebraically independent quartic invariant polynomial in the {\bf 20} of SU(1,5), \cite{Ferrara:2008ap}. There is no known anomaly. It appears  that  the prediction \cite{Kallosh:2012yy}  of the absence of the 4-loop UV divergence of ${\cal N}=5$ supergravity is confirmed in \cite{Bern:2014sna}.  The computation in  \cite{Bern:2014sna} taken together with the analysis in \cite{Kallosh:2012yy} makes it plausible that other models of $\cN\geq 5$ supergravity might be finite for the number of loops $L= \N-1$, for example in case of $\cN=8$ at 7 loops. It would also suggest that only the analysis of the scalar soft theorems due to $E_{7(7)}$ symmetry may be incomplete and the vector sector duality current conservation may need an additional study.

It may be important also to understand the recent observation in \cite{Meissner:2016onk} that the conformal anomalies proportional to a square of the Weyl tensor $C_{\mu\nu \rho\sigma} C^{\mu\nu \rho\sigma} $ as well as the ones proportional to a Gauss-Bonnet term, all cancel in ${\cal N}=5, 6, 8$ supergravities.  In conclusion, a better understanding of the complete consequences of $E_{7(7)}$ symmetry for ${\cal N}\geq 5$ supergravity would be desirable.

\subsection{D3 brane and UV completion of constrained multiplet models in string theory}

In cosmology the existence of constrained ${\cal N}=1$ superfields is helpful for inflationary model building, see for example,  \cite{Ferrara:2014kva,Ferrara:2015tyn,Carrasco:2015pla} and for a simple description of dark energy,  \cite{Bergshoeff:2015tra,Hasegawa:2015bza,Kuzenko:2015yxa,Bandos:2015xnf}. Constrained multiplets \footnote{A nilpotent multiplet with a single constraint $S^2=0$ was introduced in \cite{Casalbuoni:1988xh}, the scalar-less one, $S\, Y =0$ in \cite{Brignole:1997pe}, a  gaugino-less one of the Born-Infeld type $W^2+ S(1- {1\over 4} \bar D^2 \bar S)=0$ and $SW_\alpha=0$ in \cite{Cecotti:1986gb,Bagger:1996wp}, and finally the relaxed multiplets $S \, \bar D_{\dot \alpha} \bar H (x, \theta, \bar \theta)=0$ and the orthogonal nilpotent one $S(\Phi-\bar \Phi)=0$ in \cite{Komargodski:2009rz}.
} are associated with nonlinearly realized supersymmetry. A natural issue  in these models concerns possible role of quantum corrections, unitarity and UV completion.

It has been recognized a while ago that the nonlinear supersymmetry is associated with the physics of D-branes in string theory, see for example \cite{Hughes:1986fa,Bianchi:1991eu,Antoniadis:2004uk,Polchinski:1998rr} and references therein. But recently it become easier to address the issues of UV completion of cosmological models, when it was established in details, that constrained multiplets are present on the world-volume of the D3 brane. Namely, the nilpotent multiplet
\be
S^2(x, \theta)=0
\ee
equivalent to VA nonlinear goldstino theory provides a supersymmetric KKLT uplifting, \cite{Kallosh:2014wsa}. But also other constrained ${\cal N}=1$ superfields are present on D3 brane, \cite{Vercnocke:2016fbt,Kallosh:2016aep}, which includes `partnerless' multiplets, 
\be
S\, Y^i (x, \theta)=0\, , \qquad SW_\alpha (x, \theta)=0 \, , \qquad S \, \bar D_{\dot \alpha} \bar H (x, \theta, \bar \theta)=0
\ee
and, for example an orthogonal to the nilpotent one, which is particularly useful in the role of  an inflaton multiplet
\be
S \, B (x, \theta, \bar \theta)=0\, , \qquad B= {1\over 2i} (\Phi-\bar \Phi) \, , \qquad B^3(x, \theta, \bar \theta)=0
\ee
In view of the fact that these constrained ${\cal N}=1$ superfields are packaged on the world-volume of D3 brane, one may try to use the known information about quantum corrections to D3 brane action. The details of the action upon gauge-fixing a local fermionic $\kappa$-symmetry, are given in Appendix. The action has 16 deformed Maxwell supersymmetries and 16 VA type non-local supersymmetries.

The first-loop level logarithmic divergence to D3 action was computed in \cite{Shmakova:1999ai} using the helicity amplitudes method. The result in the 4-vector sector is 
\be
-{(2\pi \alpha')^4\over (4\pi)^2} {1\over 16 \, \epsilon} (s^2+ t^2 +u^2) \, t_8 \, F^4
\label{Marina}\ee
where $t_8 \, F^4 \equiv (t_8)_{\mu_1 \nu_1 \mu_2 \nu_2 \mu_3 \nu_3 \mu_4 \nu_4 } F^{\mu_1 \nu_1} F^{\mu_2 \nu_2} F^{\mu_3 \nu_3} F^{\mu_4 \nu_4}$ and the $t_8$ tensor $(t_8)_{\mu_1 \nu_1 \mu_2 \nu_2 \mu_3 \nu_3 \mu_4 \nu_4 }$ is defined in \cite{Schwarz:1982jn}.

A related computation was performed in \cite{DeGiovanni:1999hr}, for the ${\cal N}=2$ supersymmetric BI action using the  supergraphs in  ${\cal N}=1$ superspace. The UV divergence was also given in terms of ${\cal N}=1$ photino superfields, $W_\alpha$, $\bar W_{\dot \alpha}$ and is proportional to
\be
(s^2 + {4\over 3} t^2) W^2 \bar W^2
\label{Daniela}\ee
where one can get the 4-fermion part as well as the 4-vector part.  The 4-vector part is the one in eq. \rf{Marina}. Additional terms depend on scalars. It was noticed in \cite{DeGiovanni:1999hr}
 that such terms have been obtained in the past for scattering amplitudes of vector fields in type IIB string theory on the D3-brane  and earlier in type I open string theory \cite{Schwarz:1982jn}, where the result is
 \be
{\Gamma(s) \Gamma(t) \over \Gamma(1+s+t)} K(\epsilon^i, p^j) 
\label{Ven} \ee
 Expressions in \rf{Marina} and \rf{Daniela} correspond to the first terms in the $\alpha'$ expansion of \rf{Ven}.
 It is well known that at large $s$ and  fixed $t/s$, fixed angle, the high-energy behavior of the full string theory amplitude in \rf{Ven} is excellent. There is no unitarity problem for constrained multiplets in the context of string theory. In a situation when one can embed the constrained ${\cal N}=1$ multiplets into a package corresponding to D3 brane, one can view it as a UV completion of the models which are useful in cosmology.
 
More recent studies of  higher-derivative, higher multiplicity  perturbative as well as non-perturbative corrections to the D3-brane effective action in string theory and/or to N=4 in the Coulomb branch may be useful for the further development of these ideas, see for example 
 \cite{Green:2000ke}, 
 \cite{Lin:2015ixa}
 \cite{Bianchi:2015cta}.

\section{Summary}

The recent studies of amplitudes in theories with nonlinear symmetries, starting with \cite{Chen:2014xoa} and most recent results in \cite{He:2016vfi,Cachazo:2016njl,Carrasco:2016ldy} have the following features. The amplitudes for these models were explicitly constructed and the observation was made about the soft limits for  NLSM (nonlinear sigma models) and their supersymmetric version, VA models  and DBI-VA model \cite{Bergshoeff:2013pia}. These bosonic and fermionic models have a vanishing single soft limit, the Adler's zero, and the double soft limit is controlled by the ${G\over H}$ coset algebra.

The properties of amplitudes concerning the soft limits discovered in \cite{Chen:2014xoa,He:2016vfi,Cachazo:2016njl,Carrasco:2016ldy} may be viewed as a postdiction, based on construction of amplitudes,  rather than prediction based on symmetry. Our goal here was to use the existence of the action with nonlinear global symmetries to predict these and other features of amplitudes, i. e. to find an analog of Ward-type identities which control the amplitudes.

In the past the background field method for non-Abelian gauge theories and gravity  \cite{DeWitt:1967ub} was instrumental in describing the properties of the tree level gauge-independent S-matrix elements as well as in making predictions about the quantum corrections in these theories,  \cite{Kallosh:1974yh,Grisaru:1975ei,Goroff:1985th,vandeVen:1991gw}. Here we have generalized this method for the case of nonlinear symmetry.
Our results for tree amplitudes are the following. 

1. The sum of all n-point amplitudes $A_{i_1 \dots i_n}$ when contracted with free fields $\phi_0^i$
\be
 \sum_{n=4} ^\infty {1\over n!} \, A_{i_1 \dots i_n} \phi^{i_1}_0 \dots \phi^{i_n}_0= \big (S(\phi)- S_i \phi^i \Big ) \Big |_{\phi=\phi[\phi_0]}
\label{Tree}\ee
is also given by a functional of the background field $\phi^i$ 
\be
\phi^i [\phi_0] = \phi^i_0 + \Delta[\phi_0]
\label{phi}\ee
where $\Delta [\phi_0]$ is an infinite tree functional given in eq. \rf{tree1}. This follows from the saddle point expansion of the path
integral for the S-matrix.

This identity requires that all on-shell $n$-point amplitudes $A_{i_1 \dots i_n}$ have some properties such that the identity \rf{Tree} is valid.
A known manifestation of these properties is in soft limit of these amplitudes.

2. In general, when the symmetry transformation $\delta \phi^i$ has a constant-field independent part ${\cal R}^i_\alpha (0) \xi^\alpha$, corresponding to spontaneous breaking of symmetry, the formalism above predicts the Adler's zero for a single soft limit, in all models which have  only even amplitudes, like in VA and DBI-VA models, 
\be 
A_{i_1 \dots i_n} \phi^{i_1}_0 \dots \phi^{i_{(n-1)}}_0  {\cal R}^{i_n} _{\alpha}(0) \xi^\alpha   =0
\label{Adler}\ee
as shown in  eq. \rf{Adler}. Here we take into account that ${\cal R}^i _{\alpha}(0)= R^{i} _{\alpha}(0)$, i. e. the constant field-independent part of symmetry transformation is the same for the free field as well as for the unconstrained field in the action. When the model has even and odd amplitudes, a more general relation between $n$- and $(n-1)$-point amplitudes replaces the Adler's zero for a single soft limit and is shown in eq. \rf{n and n-1}.

3. The double-soft limit is shown here to depend universally on the commutator of two spontaneously broken symmetries. The corresponding identity has the form
\be
\Big (S_{, j i_1 i_2} {\cal R}^{i_1}_\alpha(0)  {\cal R}(0) ^{i_2} _\beta+ S_{, j i_1 } {\cal R}^{i_1}_{ \gamma} f^\gamma_{\alpha \beta} + \cdots\Big )\xi^\alpha  \xi^{'\beta} =0
\label{double}\ee
where we omitted other terms, shown in a complete form of this equation in eq. \rf{doublesoft}. The first term in \rf{double} presents the double soft limit of the background field expansion \rf{phi} as a function of free fields, when two fields have soft momenta, the second term shows that the structure constants of the commutator of two spontaneously broken symmetries enter the identity above, and therefore characterize universally the double-soft theorems.

4. Multi-soft limit is explained using the symmetry variation of the  identity used for double-soft limits. The procedure of deriving the  identity defining the $n$-soft limit of \rf{phi} is explained in the paper and suggest how to get the terms with $\cdots$ in eq. below.
\be
\Big (S_{, j i_1 \dots i_n} {\cal R}^{i_1}_{\alpha_1} (0)   \dots  {\cal R}^{i_n}_{\alpha_n}(0) + \cdots \Big )\xi^{\alpha_1}   \dots  \xi^{\alpha_n} =0
\label{multi}\ee
 The triple and higher $n$-point soft limit for nonlinear symmetries have not been studied, to the best of our knowledge.
 
 We plan to present the examples of the background field method with nonlinear symmetries in specific models in future publications. The example as to how the identities above work in VA theory will be given in \cite{Kallosh:2016lwj}.

\acknowledgments
I am grateful to I. Antoniadis, Z. Bern, M. Bianchi, J. J. Carrasco, G. Dall'Agata, B. de Wit, L. Dixon, E. Dudas, S. Ferrara, D. Freedman,  Yu-tin Huang, A. Karlsson,  S. Kuzenko, D. Murli, A. Linde, H. Nicolai, G. Pradisi, R. Roiban, A. Van Proeyen, B. Vercnocke, C. Wen,  T. Wrase and F. Zwirner for enlightening discussions. 
I am grateful to S. He and F. Cachazo for the discussions of  the results of their work \cite{He:2016vfi,Cachazo:2016njl}, and to the participants of the  workshop `Supergravity: what next?' in GGI in Firenze for the stimulating atmosphere when this work was completed.
This work  is supported by SITP and by the NSF Grant PHY-1316699.
\appendix
\section{Examples of models with nonlinear (super)symmetry}
An example of Volkov-Akulov model with 4 nonlinear supersymmetries
\be
{\cal L}_{VA}(\lambda, \bar \lambda) = -{1\over \kappa^2} \,  \det A
\label{VAor}\ee
where $A^\nu_\mu\equiv \delta ^\nu_\mu +i\kappa^2 (\lambda \partial_\mu  \sigma^\nu \bar \lambda - \partial_\mu \lambda \sigma^\nu \bar \lambda )  $. The nonlinear supersymmetry of this action is shown in eq. \rf{susy}.
The details of this symmetry together with other bosonic symmetries and the corresponding Noether currents and their algebra can be found in 
 \cite{Clark:1988es}.

Another interesting example of the nonlinear supersymmetry is the DBI-VA model with 16 Maxwell-type deformed supersymmetries and 16 VA type nonlinear supersymmetries. 
The gauge-fixed D3 brane model  presented in \cite{Bergshoeff:2013pia} is a  d=4  theory which has 16 deformed Maxwell type linear supersymmetries, it is an ${\cal N}=4$ Maxwell model at the linear level,  and 16 nonlinear Volkov-Akulov type supersymmetries. 
The fields on  D3-brane
  are: a  gauge field $A_\mu$, 6 scalars $\phi^I$ and a 16-component spinor $\lambda$. The geometric action is
\begin{equation}
\label{actionBI} S =  -\frac{1}{\alpha^2} \int d^{4} \sigma\,
\sqrt{- \det (G_{\mu\nu} + \alpha {\cal F}_{\mu\nu})}  \,,
\end{equation}
where $G_{\mu\nu}$ and ${\cal F}_{\mu\nu}$ are given by a fermion-dependent pull-back to the world-volume of the ten-dimensional  geometry and a 2-form:
\begin{eqnarray}
G_{\mu\nu} &=& \eta_{m n} \Pi_\mu^{m}
\Pi_\nu^{n}= \eta_{m' n'} \Pi_\mu^{m'}
\Pi_\nu^{n'}+\delta_{IJ} \Pi_\mu^I\Pi_\nu^J \,, \quad  m= 0,1\cdots 9, \quad m'=0,1,2,3 \,, \quad I=1,...,6\,,\nonumber \\[.2truecm]
\Pi_\mu^{m'} &=& \delta^{m'}_{\mu}
-\alpha^2\bar\lambda \Gamma^{m'}
\partial_\mu \lambda \ , \qquad \Pi_\mu^I = \partial_\mu\phi^I
-\alpha^2\bar\lambda \Gamma^I
\partial_\mu \lambda\,, \qquad {\cal F}_{\mu\nu} \equiv F_{\mu\nu} - b_{\mu\nu} \,,\nonumber\\
 b_{\mu\nu} &=&2\alpha\bar{\lambda}\Gamma_{[\mu }\partial_{\nu]}\lambda
-2\alpha\bar{\lambda}\Gamma_{I}\partial_{[\mu}\lambda\partial_{\nu]}
\phi^I =-2\alpha\bar{\lambda}\Gamma_{m'}\partial_{[\mu}\lambda\, \Pi_{\nu]}^{m'}
-2\alpha\bar{\lambda}\Gamma_{I}\partial_{[\mu}\lambda \, \Pi_{\nu]}
^I  \, .
\label{defGPicFD3}
\end{eqnarray}
The 32-component global supersymmetry of the action consists of  
 16
$\epsilon$-supersymmetries corresponding to a deformation of the original
16 supersymmetries of the $\cN=4$, $d=4$ Maxwell multiplet
\begin{eqnarray}
\delta_\epsilon  \phi^I &=& \ft12\alpha\bar\lambda\Gamma^I\left[\mathbbm{1}+\beta\right] \epsilon
+\xi^\mu_\epsilon \partial_\mu \phi^I\,, \nonumber \\
\delta_\epsilon  \lambda &=&- \ft1{2\alpha}\left[\mathbbm{1}  -
\beta\right]\epsilon
+\xi^{\mu}_\epsilon \partial_{\mu}\lambda\,, \nonumber \\
 \delta_\epsilon  A_\mu&=&-\ft12\bar \lambda\big(
 \Gamma_\mu + \Gamma_I\partial_\mu\phi^I\big) \left[\mathbbm{1} + \beta\right]\epsilon
+\ft12 \alpha^2 \bar \lambda \Gamma_m\left[ \tfrac{1}{3}\mathbbm{1}  +\beta\right] \epsilon
 \bar\lambda \Gamma^m \partial_\mu \lambda+\xi_\epsilon ^\rho F_{\rho\mu} \,.
 \label{epsilontransMaxw}
\end{eqnarray}
The other 16
$\zeta$-supersymmetries correspond to VA-type supersymmetries.
\begin{eqnarray} \label{decompositions1}
\delta_{\zeta} \phi^I &=& -\alpha{\bar\lambda}\Gamma^I\zeta
+\xi ^{\mu}_\zeta \partial_\mu \phi^I\,, \\
\delta_{\zeta} \lambda &=& \alpha^{-1}\zeta +\xi ^{\mu}_\zeta\partial_{\mu}\lambda\,, \\
\label{decompositions2}
 \delta_{\zeta} A_\mu&=&\bar\lambda \big(
 \Gamma_\mu + \Gamma_I\partial_\mu\phi^I\big)\zeta
+ \xi ^{\rho}_\zeta F_{\rho\mu}
-\tfrac{1}{3} \alpha^2 \bar\lambda
\Gamma_m \zeta \bar\lambda \Gamma^m \partial_\mu \lambda \,,
\label{decompositions3}
\end{eqnarray}
where
$
\xi^{\mu}_{\epsilon} \equiv
-\tfrac{1}{2}\alpha\bar{\lambda}\Gamma^{\mu}\left(\mathbbm{1} +
\beta\right)\epsilon\,,\xi ^\mu _\zeta = \alpha\bar\lambda \Gamma ^\mu \zeta 
$, $\Gamma_{(0)}^{D3} =
\frac{1}{4!\sqrt{|G|}}\varepsilon^{\mu_1\dots \mu_{4}}\hat \Gamma_{\mu_1\dots
\mu_{4}}\,,\Gamma _*^{(3)}=-\rmi\Gamma ^0\Gamma ^1\Gamma ^2\Gamma ^3\,,
$
$  \beta= -i {\cal G} \sum^{2}_{k=0} \frac{\alpha^k}{2^k k!} \hat \Gamma^{\mu_1
\nu_1 \cdots \mu_k \nu_k }\mathcal{F}_{\mu_1 \nu_1}\cdots
\mathcal{F}_{\mu_k \nu_k}\Gamma _{(0)}^{D3}\Gamma _*^{(3)}$, ${\cal G}=\frac{\sqrt{\left|G\right|}}{\sqrt{\left| G + \alpha \cal{F}\right|}}=
\left[ \det\left( \delta _\mu {}^\nu +\alpha {\cal F}_{\mu \rho }G^{\rho \nu }\right) \right] ^{-1/2}\,,
$
The action also has a global shift symmetry
\be
\delta\phi^I= a^I
\ee
This symmetry is a surviving part of the Poincar{\'e} translation symmetry
in $d=10$, $\phi^I=X^{I}\rightarrow X^I+ a^I$ for $I=1,...,6$. Another symmetry is the local $U(1)$ symmetry of the gauge-field $A_\mu$.

The linearized action (\ref{actionBI}) and the linearized susy $\epsilon $-rules in \rf{epsilontransMaxw} were shown in  \cite{Bergshoeff:2013pia} to give an $\cN=4$ Maxwell action,
namely
\begin{equation}
S_{\rm Maxw}^{\cN=4}= \int d^4 \sigma \Big (-\ft14 F_{\mu\nu} F^{\mu\nu }  + 2\bar
\psi_a \slashed{\partial} \psi^a -\ft18\partial _\mu\varphi_{ab}
\partial^\mu \varphi^{ab}\Big ) \,,
\label{quadr}
\end{equation}
where  the $d=4$ left-handed chiral spinor is
assigned to the fundamental representation of $SU(4)$, and carries an
upper $SU(4)$ index; right-handed components will then transform
according to the conjugate representation,  and have a lower index. The scalars $\varphi_{ab}$ are self-dual, $\varphi^{ab} = -{1\over 2} \epsilon^{abcd} \varphi_{cd}$.
The action (\ref{quadr}) is invariant under the 4 linear $\epsilon^a$ supersymmetries.
All fields of this linear $\cN=4$ vector multiplet, a vector, 4 spinors and 6 scalars  are linear  $\epsilon$-partners. However,  at the  nonlinear level  these fields belong to constrained  multiplets,  \cite{Vercnocke:2016fbt,Kallosh:2016aep}.

\bibliographystyle{JHEP}
\bibliography{refs}

\end{document}